\documentstyle[11pt]{article}
\begin{document}

\newtheorem{theorem}{Theorem}
\renewcommand{\theequation}{\thesection.\arabic{equation}}
\catcode`\@=11
\@addtoreset{equation}{section}

\title{Limit Set of Trajectories of 
the Coupled Viscous Burgers' Equations 
\footnote{Address for
correspondence: Dr. J. Duan, Department of Mathematical Sciences,
Clemson University, Clemson, South Carolina 29634, USA.
E-mail address: duan@math.clemson.edu;
Tel: (864)656-2730; Fax: (864)656-5230.}  }
\author{}
\date{ July 25, 1996  }
\maketitle

\begin{center}
Janpou Nee\\Institute of  Mathematics,
Academia Sinica\\Taipei, 11529, Taiwan, Republic of China.\\
        and \\
Jinqiao Duan
\\Department of Mathematical Sciences\\
Clemson University\\
Clemson, South Carolina 29634, USA.
\end{center}

\begin{abstract}
In this letter, a coupled system of viscous Burgers' equations with
zero Dirichlet boundary conditions and appropriate initial data is
considered. For the well-known single viscous Burgers' equation with
zero Dirichlet boundary conditions, the zero equilibrium is the unique
global exponential point attractor.  A similar property is shown for
the coupled Burgers' equations, i.e., trajectories starting with
initial data which is not too large approach the zero equilibrium as
time goes to infinity.  This ``approaching" or convergence is not
necessarily exponentially fast, unlike the single viscous Burgers'
equation.

\vspace{1cm}


\end{abstract}

\section{Introduction}

	We consider the following coupled Burgers' 
equations
\begin{equation}
    \cases{
       u_t = u_{xx}-uu_x - a(uv)_x,\cr
       v_t = v_{xx}-vv_x - b(uv)_x,\cr}
    \label{eqn}
\end{equation}
together with Dirichlet boundary conditions
\begin{equation}
    \cases{
       u(0,t)=u(1,t)=0,\cr
       v(0,t)=v(1,t)=0,\cr}
    \label{bc}
\end{equation}
and appropriate initial conditions
\begin{equation}
    \cases{
        u(x,0)=u_0(x),\cr
        v(x,0)=v_0(x),\cr}
    \label{ini}
\end{equation}
for $0<x<1, t>0$. Here $a, b$ are constants.

This coupled system, derived by Esipov \cite{Esipov}, is a simple model
of sedimentation or evolution of scaled volume concentrations of two
kinds of particles in fluid suspensions or colloids, under the effect
of gravity.  The constants $a, b$ depend on the system parameters such
as the P\'{e}clet number, the Stokes velocity of particles due to
gravity, and the Brownian diffusivity. In \cite{Esipov} Esipov reported
numerical simulations for (\ref{eqn})-(\ref{ini}) and compared the
results with experimental data.

In this letter, we consider dynamical aspect of the coupled system.  We
show that trajectories (orbits of solutions) of this coupled system
approach the zero equilibrium as $t \rightarrow +\infty$ when the
initial data $u(x,0)$, $v(x,0)$, is not too large in some Sobolev norm.
That is, the $\omega$-limit set of these trajectories is the zero
equilibrium.

\section{Single viscous Burgers' equation}

We first recall an interesting property about $\omega$-limit set
for the well-known
(single) viscous Burgers' equation (\cite{Burgers}), i.e. equations in
(\ref{eqn}) without the nonlinear coupling term
\begin{equation}
w_t = w_{xx}-ww_x,
\label{burgers}
\end{equation}
with boundary and initial conditions
\begin{equation}
    \cases{
	w(0,t)=w(1,t)=0,\cr
        w(x,0)=w_0(x).\cr}
\end{equation}
The interest in the Burgers' equation arises because it is a simple one
dimensional analog of the Navier-Stokes equation.  The importance of
the Burgers' equation is due to the nonlinear convection term $uu_x$.

In the following, $L^2(0,1)$, $L^{\infty}(0,1)$,
$H_0^1(0,1)$ and $H_0^2(0,1)$ are the usual Sobolev
spaces, while $C(0,1)$ is the space of continuous functions.
We denote by $||\cdot ||$ the usual $L^2(0,1)$ norm.
All integrals $\int$ are with respect to 
$x\in [0,1]$, unless specified otherwise.

Multiplying the equation (\ref{burgers}) by $w$ and integrating
over $x\in [0,1]$, we get
\begin{equation}
\frac12 \frac{d}{dt}\|w\|^2 = - \|w_x\|^2 \leq -\|w\|^2,
\end{equation}
where we have used the Poincar\'{e} inequality
\begin{equation}
\int_0^1 w^2dx \leq \int_0^1 w_x^2dx,
\end{equation}
in the last step.
Thus by the Gronwall inequality (\cite{Temam}, p.88), we further have
\begin{equation}
\|w(x,t)\|^2 \leq \|w(x,0)\|^2 e^{-2t}, \;\;\; t>0.
\end{equation}
This means that all trajectories converge in the $L^2(0,1)$- norm to
the zero equilibrium exponentially fast, i.e., the $\omega$-limit set
(\cite{Hale}) of every trajectory is the zero equilibrium. The zero
equilibrium is the global point attractor.  See \cite{Ly} and
\cite{Burns} for further results in this regard.  This property also
holds for higher dimensional viscous Burgers type convection-diffusion
equations (\cite{Hill}).

\section{$\omega$-limit set of the coupled viscous Burgers' equations}

In this section, we show that trajectories of the coupled viscous
Burgers' equations (\ref{eqn})-(\ref{ini}), whose initial data is not
too large in $H^1_0(0,1)$-norm, converge to the zero equilibrium in the
$\max$-norm.  This convergence, though, is not necessarily
exponentially fast, unlike the situation for the single viscous
Burgers' equation.

  Let
\begin{eqnarray}
    I(t) =\frac{1}{2}(||u||^2+||v||^2),\label{ene} \\
    J(t)=\frac{1}{2}(||u_x||^2+||v_x||^2),\label{ener}  \\
    K(t)=\frac{1}{2}(||u_{xx}||^2+||v_{xx}||^2).\label{ener2}
\end{eqnarray}

For initial data $u(x,0), v(x,0) \in H^1_0(0,1)$,
local (-in-time) existence and uniqueness of classical solutions to
(\ref{eqn})-(\ref{ini}) can be easily shown
by the usual semigroup method; cf. \cite{Henry}, Theorems 3.3.3, 3.3.4
and 3.5.2.  Global existence and uniqueness of classical solutions then
follows once we show a priori that the solutions do not become
unbounded in $H^1_0(0,1)$-norm at finite positive time. Moreover, the
solutions, whenever they exist, are in $H^k_0(0,1), k=1,2,\cdots$; (cf.
\cite{Henry}, p.73).  Note that $I(t)\leq J(t)\leq K(t)$ whenever $u,
v$ exist, by the Poincar\'{e} inequality.

We will show that $J(t) \rightarrow 0$
as $t \rightarrow 0$, when $J(0)$ is bounded by some
constants depending only on the system parameters $a,b$ in (\ref{eqn}).

Multiplying the first and second
equation in (\ref{eqn}) by $-u_{xx}$ and $-v_{xx}$, respectively,
adding the two resulting equations and then integrating over
$x\in[0,1]$, we obtain
\begin{eqnarray}\label{dj}
    \frac{d J}{dt}
    & =  & \int-u_{xx}^2+uu_xu_{xx}+a(uv)_xu_{xx}
	   -v_{xx}^2+vv_xv_{xx}+b(uv)_xu_{xx}	\nonumber\\
    & =  & \int-u_{xx}^2-\frac{u_x^3}2
	   -v_{xx}^2-\frac{v_x^3}2+ (a+b)(uv)_x(u_{xx}+v_{xx})\nonumber\\
    &\leq& \int-u_{xx}^2-\frac{u_x^3}2-v_{xx}^2-\frac{v_x^3}2\nonumber\\
    &	 & +(|a|+|b|)(\int(u_xv+uv_x)^2)^{\frac12}
	   (\int(u_{xx}+v_{xx})^2)^{\frac12}\nonumber\\
    &\leq& -2K-\int\frac{u_x^3}2 - \int\frac{v_x^3}2
	   +(|a|+|b|)(\int 2(u_x^2v^2+u^2v_x^2))^{\frac12}
	   (2K)^{\frac12}\nonumber\\
    &\leq& -2K-\int\frac{u_x^3}2-\int\frac{v_x^3}2\nonumber\\
    &	 & +2(|a|+|b|)\{\sqrt{\int u_x^4\int v^4}
	   +\sqrt{\int u^4\int v_x^4}\}^{\frac12}K^{\frac12}.
\end{eqnarray}
Now we further estimate the right hand side of (\ref{dj}) term by term. 

From the fact that
\[ u^2 = 2\int_0^x uu_xdx \leq 2 ||u||\cdot ||u_x||, \]
for $u\in H^1_0(0,1)$, we get the so-called Agmon inequality
\begin{equation}
||u||_{\infty}^2 \leq 2 ||u||\cdot ||u_x||,
\end{equation}
where $||u||_{\infty}$ is the
$L^{\infty}$-norm. 

Observe, using the Cauchy-Schwarz inequality  
(\cite{Renardy}, p.183) and the above Agmon inequality
\begin{eqnarray}\label{ux3}
	\int u_x^3 dx 
& =  & \int u_x^2 du = -2\int uu_xu_{xx} dx \nonumber\\
&\leq& 2\|u\|_{\infty} (\int u_x^2)^{\frac12} (\int u_{xx}^2)^{\frac12}
						\nonumber\\
&\leq& 2\sqrt{2}(\int u^2)^{\frac14}(\int u_x^2)^{\frac14}
	(\int u_x^2)^{\frac12} (\int u_{xx}^2)^{\frac12}
						\nonumber\\
&=   & 2\sqrt{2}(\int u^2)^{\frac14}(\int u_x^2)^{\frac34}
	(\int u_{xx}^2)^{\frac12}
                                                \nonumber\\
&\leq& 8 I^{\frac14}J^{\frac34}K^{\frac12}      \nonumber\\
&\leq& 8 JK^{\frac12}
\end{eqnarray} 
In the last step, we have used the fact that $I(t)\leq J(t)$.
The same inequality holds for $\int v_x^3dx$.

Similarly,
\begin{eqnarray}
        \int u_x^4 dx
& =  & \int u_x^3 du = -3\int uu_x^2 u_{xx} dx  \nonumber\\
&\leq& 3\|u\|_{\infty}(\int u_x^4)^{\frac12} (\int u_{xx}^2)^{\frac12}
                                                \nonumber\\
&\leq& 3\sqrt{2}(\int u^2)^{\frac14}(\int u_x^2)^{\frac14}
	(\int u_x^4)^{\frac12} (\int u_{xx}^2)^{\frac12}.
\end{eqnarray}
Thus
\begin{eqnarray} \label{ux4}
        (\int u_x^4 dx)^{\frac12}
&\leq& 3\sqrt{2}(\int u^2)^{\frac14}(\int u_x^2)^{\frac14}
		(\int u_{xx}^2)^{\frac12}
					\nonumber\\
&\leq& 6\sqrt{2} I^{\frac14}J^{\frac14}K^{\frac12} \nonumber\\
&\leq& 6\sqrt{2} J^{\frac12}K^{\frac12}.
\end{eqnarray}
The same inequality holds for $\int v_x^4dx$.

Moreover, note that
    \[u^4=(\int_0^x 2uu_xdx)^2
    \leq 4\int u^2dx\int u_x^2dx.\]
So we have
\begin{eqnarray}\label{u4}
    \int u^4dx\leq 4\int u^2dx\int u_x^2dx \leq 16IJ \leq 16J^2.
\end{eqnarray}
The same estimate holds for $\int v^4dx$.

    Substituting (\ref{ux3}), (\ref{ux4}) and (\ref{u4}) into (\ref{dj}), 
and using the {\em Young's inequality} (\cite{Temam}, p.108), we
finally get
\begin{eqnarray}
    \frac{d J}{dt}
    &\leq& -2K+ 8 JK^{\frac12} +8\sqrt{3}2^{\frac14}
			(|a|+|b|)J^{\frac34}K^{\frac34}
					 \nonumber\\
    &\leq& -2K+ \frac{\epsilon}2K+\frac1{2\epsilon}64J^2
		+\frac{3\epsilon}4 K 
		+\frac1{4\epsilon^3}	
		[8\sqrt{3}2^{\frac14}(|a|+|b|)]^4 J^3,
\end{eqnarray}
for $\epsilon >0$.
Taking $\epsilon=\frac45$ and noting that $J(t)\leq K(t)$, we now have
\begin{eqnarray} \label{djt}
\frac{d J}{dt}
    &\leq& -K + 40J^2 + 36000 (|a|+|b|)^4 J^3   \nonumber\\
    &\leq& -J + 40J^2 + 36000 (|a|+|b|)^4 J^3.
\end{eqnarray} 
The comparison equation 
\begin{eqnarray} \label{comp}
\frac{dJ}{dt} = -J + 40J^2 + 36000 (|a|+|b|)^4 J^3 \equiv f(J)
\end{eqnarray}
has fixed points $0$ and 
\[ J_{+}= \frac{-1+\sqrt{1+9000(|a|+|b|)^4}}{1800(|a|+|b|)^4}. \]
The third fixed point is negative and is discarded since $J(t)$ is always
non-negative by definition (\ref{ener}).
We calculate $f'(0) <0$ and $f'(J_{+})>0$. So fixed point $0$ is 
stable while $J_{+}$ is unstable; see \cite{Smale}, p.187 or
\cite{Wiggins}, p.8.
That is, for the comparison equation (\ref{comp}),
$J(t) \rightarrow 0$ if $J(0) < J_{+}$, while 
$J(t) \rightarrow \infty$ if $J(0) > J_{+}$.
Due to the standard comparison result for
ordinary differential inequalities and equations (\cite{Walter}, P.69),
the $J(t)$ satisfying
the differential inequality (\ref{djt}) also approach zero
as $t$ goes to infinity, when $J(0) < J_{+}$.

We have thus shown that, if the initial data
satisfies $\frac12 (\|u_0'(x)\|^2+\|v_0'(x)\|^2) < J_{+}$, the corresponding
classical solutions and hence
trajectories exist for all $t>0$, since the $H^1_0(0,1)$-norm,
i.e., $J(t)$, in this case
is bounded. Moreover these trajectories approach the zero equilibrium
as $t \rightarrow \infty$.
The zero equilibrium is the $\omega$-limit set of these trajectories.
We remark that this convergence is not necessarily exponentially fast,
unlike the single Burgers' equation.
If $\frac12 (\|u_0'(x)\|^2+\|v_0'(x)\|^2) > J_{+}$, however, we cannot conclude
any thing about the corresponding trajectories,
based on the above dynamical system style analysis.

Note that $\|u_x\| $, $\|v_x\|$ are actually $H^1_0(0,1)$-norms of $u$
and $v$ due to the Poincar\'{e} inequality, and note also that
$H^1_0(0,1)$ is embeded in $C(0,1)$. So the above convergence of
trajectories holds in the $\max$-norm in $C(0,1)$.

Therefore we obtain the following theorem
\begin{theorem}
Assume that $u_0(x), v_0(x) $ are in $H^1_0(0,1)$ and satisfy
\[ 
\frac12 (\|u_0'(x)\|^2+\|v_0'(x)\|^2) 
	< \frac{-1+\sqrt{1+45(|a|+|b|)^4}}{225(|a|+|b|)^4}. 
\]
Then the global unique classical solutions of the coupled
system (\ref{eqn})-(\ref{bc})-(\ref{ini}) exist, and
the corresponding trajectories (orbits of solutions) approach 
the zero equilibrium in $\max$-norm.
That is, the zero equilibrium is the $\omega$-limit 
set of these trajectories.
\end{theorem}

\medskip
 
{\bf Acknowledgement.}  We would like to thank Vince Ervin for useful
comments of this work.

\end{document}